\documentclass[preprint,12pt]{elsarticle}

\usepackage{notoccite}

\usepackage{amssymb}
\usepackage{amsmath}
\usepackage{textcomp}
\usepackage{float}
\usepackage{caption}

\usepackage{multirow}
\usepackage{booktabs}
\usepackage{tikz}
\usetikzlibrary{shapes,arrows}
\usepackage{gensymb}

\tikzstyle{line} = [thick,->,>=stealth]

\tikzstyle{decision} = [diamond, draw, fill=green!30, 
    text width=10em, text badly centered, inner sep=0pt]
\tikzstyle{process} = [rectangle, draw, fill=orange!30, 
    text width=10em, text centered, rounded corners, minimum height=5em]
\tikzstyle{input} = [rectangle, draw, fill=blue!30, 
    text width=10em, text centered, rounded corners, minimum height=5em]
\tikzstyle{block} = [rectangle, draw, fill=none, 
    text width=10em, text centered, rounded corners, minimum height=5em]
\tikzstyle{line} = [draw, -latex']

\usepackage{adjustbox}
\usepackage[justification=justified]{caption}

\journal{Materials Characterization}

\begin{document}

\begin{frontmatter}


\title{An image-driven machine learning approach to kinetic modeling of a discontinuous precipitation reaction}

\author[PNNL]{Elizabeth Kautz \corref{Liz} \fnref{label1}}
 \address[PNNL]{Pacific Northwest National Laboratory, 902 Battelle Boulevard, P.O. Box 999, Richland, WA 99352, United States}
 \ead{elizabeth.kautz@pnnl.gov}
 
 \author[rpics]{Wufei Ma \fnref{label1}}
 \ead{maw3@rpi.edu}
 
 
\author[PNNL]{Saumyadeep Jana}
 \ead{saumyadeep.jana@pnnl.gov}

\author[PNNL]{Arun Devaraj}
\ead{arun.devaraj@pnnl.gov}
 
\author[PNNL]{Vineet Joshi}
\ead{vineet.joshi@pnnl.gov}

\author[rpics]{B\"{u}lent Yener}
\address[rpics]{Computer Science Department, Rensselaer Polytechnic Institute, Troy, NY 12180, USA }
\ead{yener@cs.rpi.edu}
 
 \author[rpimse]{Daniel Lewis \corref{Dan}}
 \ead{lewisd2@rpi.edu}
 \address[rpimse]{Materials Science and Engineering Department, Rensselaer Polytechnic Institute, Troy, NY 12180, USA }
 
\cortext[Liz]{Corresponding Author}
\fntext[label1]{E.K. (Author One) and W.M. (Author Two) contributed equally to this work.}

\begin{abstract}

Micrograph quantification is an essential component of several materials science studies. 
Machine learning methods, in particular convolutional neural networks, have previously demonstrated performance in image recognition tasks across several disciplines (e.g. materials science, medical imaging, facial recognition). Here, we apply these well-established methods to develop an approach to microstructure quantification for kinetic modeling of a discontinuous precipitation reaction in a case study on the uranium-molybdenum system. Prediction of material processing history based on image data (classification), calculation of area fraction of phases present in the micrographs (segmentation), and kinetic modeling from segmentation results were performed. 
Results indicate that convolutional neural networks represent microstructure image data well, and segmentation using the k-means clustering algorithm yields results that agree well with manually annotated images.  Classification accuracies of original and segmented images are both 94\% for a 5-class classification problem. Kinetic modeling results agree well with previously reported data using manual thresholding. The image quantification and kinetic modeling approach developed and presented here aims to reduce researcher bias introduced into the characterization process, and allows for leveraging information in limited image data sets.

\end{abstract}
\end{frontmatter}
\section{Introduction}
\label{intro}

Phase transformations and microstructural evolution in material systems are often studied extensively using imaging techniques, but a gap in such studies is a reproducible, quantitative description of microstructure image data that can then be linked to variables of interest (e.g. processing parameters, material chemistry). Typically, quantitatively linking image data to processing history relies on significant domain knowledge and manual or semi-automated image analysis. Such an approach to image analysis can be biased and inefficient. To address this need for an improved approach to microstructure quantification for materials science studies, we test the applicability of machine learning methods for studying the kinetics of a discontinuous precipitation reaction. Such a phase transformation has previously been observed in several metal alloys including  Mg-Al \cite{DP_MgAl}, Fe-Cr \cite{Srinivas1997,Sennour2004}, Cu-Ni-Sn \cite{Alili2008}, and U-Mo \cite{Devaraj2018_Scripta}. Here, we perform a case study on the discontinuous precipitation reaction observed in uranium alloyed with 10 wt\% Mo (U-10Mo).
 
U-10Mo is currently under investigation as a candidate low enriched uranium (LEU) fuel system in order to replace highly enriched uranium (HEU) fuels currently used in several research and radioisotope production facilities worldwide. Replacement of HEU with a LEU fuel system will reduce proliferation concerns associated with continued operation with HEU fuels \cite{Berghe2014,Neogy2017,Ugajin1998,Snelgrove1997,Meyer2002,Kim2011,IAEA1996}. 

A monolithic, plate-type design has recently received significant attention for this new LEU fuel system due to the increased U density achievable in comparison to the dispersion design. High U density is needed to meet reactor performance requirements of U.S. High Performance Research Reactors (HPRRs) \cite{Burkes2015a, Burkes2015}. However, the U-10Mo monolithic fuel system requires several thermo-mechanical processing steps in order to create the final fuel form. Microstructure generated as a function of fuel processing must therefore be well-characterized in order to develop microstructure-processing-property relationships necessary for nuclear fuel qualification \cite{Jana2017, Meyer2014}. 
One particularly important aspect of on-going U-10Mo microstructure characterization efforts is understanding the mechanisms and kinetics of phase transformations observed during temperatures typical of hot isostatic pressing (HIP).  Prior work has reported  a phase transformation of the U-10Mo matrix during annealing treatments in the temperature range expected for HIP (approximately 450-525\degree C) \cite{Jana2017,Devaraj2018_Scripta}. The phase transformation pathway has been detailed in our prior work, and involves both discontinuous and continuous precipitation reaction types, as follows \cite{Devaraj2018_Scripta}: 

\begin{centering}
$\gamma$-UMo $\rightarrow$ $\alpha$-U + Mo-enriched $\gamma$-UMo (discontinuous) \\
$\gamma$-UMo $\rightarrow$ $\gamma$'-U${_2}$Mo  (continuous) \\
\end{centering}

\begin{figure}
  \centering
  \includegraphics[scale=0.6]{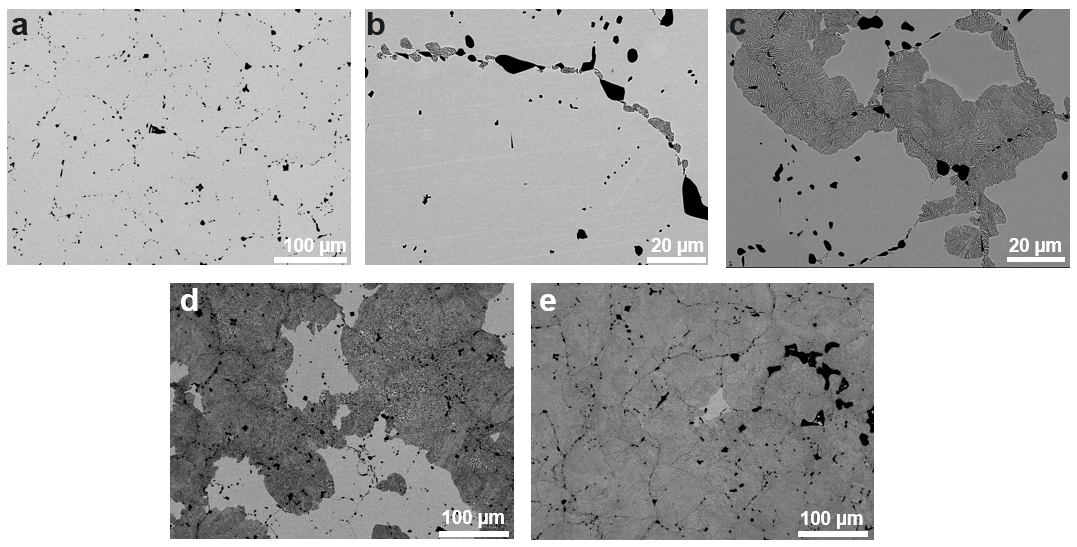}
  \caption{Transformation sequence observed in an as-cast, homogenized U-10Mo alloy that is then subjected to a series of sub-eutectoid annealing treatments from 1-100 hours at 500\degree C. Each image is of a U-10Mo sample that was cast and homogenized at 900\degree C for 48 hours. The micrograph in (a) is the as-cast and homogenized alloy. The subsequent micrographs are of samples that were annealed at 500\degree C for the following times: (b) 1 hour, (c) 10 hours, (d) 50 hours, (e) 100 hours. Micrograph magnifications were selected in order to clearly show the microstructural features of interest in this particular study (matrix, uranium carbide inclusions, and lamellar transformation products).} \label{fig:sequence}
\end{figure}

Figure \ref{fig:sequence} illustrates the microstructure evolution of U-10Mo after casting and homogenization, and subsequent sub-eutectoid annealing treatments at 500\degree C for 1 to 100 hours. The micrographs shown here demonstrate the progression of the discontinuous precipitation reaction, where the lamellar transformation products consume prior $\gamma$-UMo matrix grains with increasing time at 500\degree C. 

These lamellar transformation products (also referred to here as discontinuous precipitation products) formed during this discontinuous precipitation reaction have significant implications to fuel performance; the $\alpha$-U phase has poor swelling behavior during neutron irradiation \cite{Roberts1956,Pugh1961,Hudson1962}. Thus, the formation of any $\alpha$-U must be minimized in the fuel fabrication process, because any non-uniform swelling of fuel in an operating research reactor could impact mechanical integrity of fuel elements. Improving predictive capabilities related to the kinetics of this lamellar phase transformation is needed to predict the extent of this transformation, and improve materials process design. 

\begin{figure}
  \centering
  \includegraphics[scale=0.6]{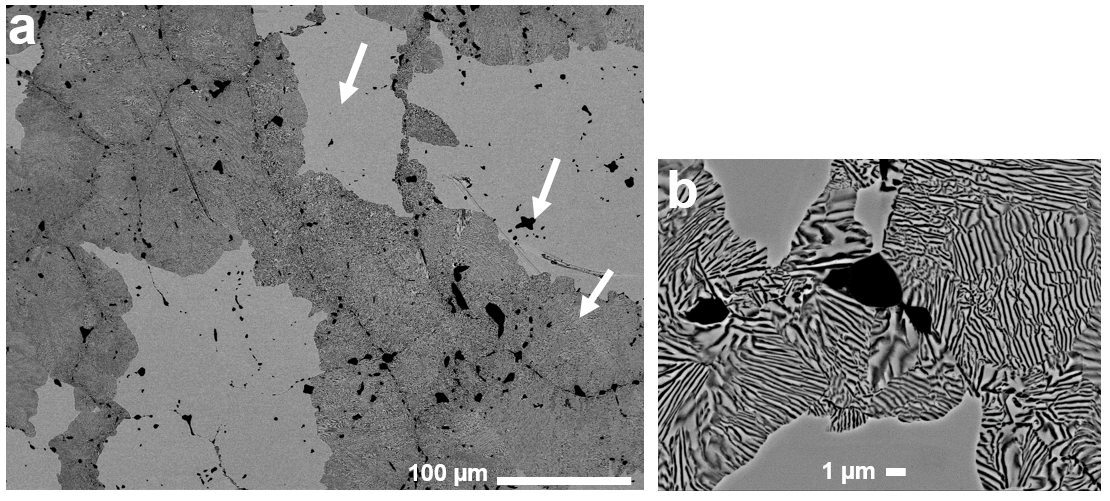}
  \caption{Microstructural features of interest in the U-10Mo system. The main features of interest are indicated by white arrows in (a), and include: the $\gamma$-UMo matrix phase, UC inclusions, and lamellar transformation products, where a magnified view of these lamellae are provided in (b).} \label{fig:microstructural_features}
\end{figure}

In addition to these lamellar transformation products, a matrix phase and inclusions are also visible in micrographs. The specific microstructural features of interest in this system are clearly labeled in Figure \ref{fig:microstructural_features} and include the following: (1) the $\gamma$-UMo matrix phase, (2) uranium carbide (UC) inclusions, and (3) lamellar transformation products. At lower magnifications (for example, in Figure \ref{fig:microstructural_features}(a)), the lamellae appear as a continuous darker gray-scale region, but at higher magnifications the alternating phases are more clearly shown, as in Figure \ref{fig:microstructural_features}(b). It is also noted here that the black/dark regions are referred to here as a UC phase, but prior work has indicated that these inclusions can also include a phase that is Si-rich.  Thus, the microstructural feature referred to in this work as UC can either be UC or a U$_2$MoSi$_2$C phase \cite{Kovarik_SiRichPhase}. For the purposes of image analysis in this study, distinguishing between UC and U$_2$MoSi$_2$C phases is not needed, but the authors acknowledge that different inclusion phases exist in this material system. 

Several methods have been used to quantitatively analyze the lamellar transformation products formed during sub-eutectoid annealing treatments, including: X-ray diffraction, differential scanning calorimetry, and optical and scanning electron microscopy. Analysis of micrographs is the only way to detect (and quantify) the extent of this phase transformation when it is less than 20\% \cite{Jana2017}. Thus, to build a predictive model relating processing parameters (time, temperature) to microstructure, image data (specifically from a scanning electron microscope, SEM) must be used.
While American Society for Testing and Materials (ASTM) standards exist for quantifying some microstructural features (e.g. average grain size measurement and nodule count in ductile iron \cite{ASTM_grainsize,ASTM_noduleCount}) these standards are limited in terms of applicability to very few material systems and microstructures. Existing standards are insufficient for scientists looking to develop quantitative microstructure-processing-properties relationships rapidly, without significant development time for different material systems. More complex microstructural features, such as impurity phase precipitates, inclusions, voids, dendrites, and lamellae are typically quantitatively described using methods specifically developed for specific images and systems. 


Recently, the use of machine learning methods has gained popularity for use in a wide range of materials science and engineering applications \cite{Ling2017,Chowdhury2016, DeCost2018,DeCost2015,DeCost2017,Kautz2019,Johns2017,Butler2018,Nikolaev2016, Ye2018, Oses2018, Draxl2018, Menon2017, Plata2017,Ward2018, Kalinin2016,Fang2009, Yang2018, Ballard2017, Chmiela2018, Jose2012,Bartok2017}.  A rapidly growing area in machine learning in materials science is in image data (micrograph) quantification. Prior work has demonstrated success of convolutional neural networks (CNNs) in microstructure recognition tasks without significant development time, and state-of-the-art performance for a wide range of microstructures (e.g. forged titanium, perlitic steel, metal powder, ceramics) \cite{DeCost2015,DeCost2017,DeCost2018,Ling2017,Kondo2017}. Taking this micrograph representation and linking it to material properties, kinetics, or performance has received less attention today, but is the natural next step beyond the simpler recognition task. 

Additionally, the application of machine learning methods to large image data sets, such as those available through ImageNet is routinely done \cite{Krizhevsky2012,Deng2009}. The ImageNet database includes over 14 million natural images that can be used for training and testing of machine learning models. Application of machine learning methods to limited data sets is still a frontier in the machine learning community \cite{Butler2018, Lake2015,Kondo2017,Kautz2019,Mace2018}, and is of interest to materials science data analysis problems, where only limited or historic data sets are available, and the cost/time associated with obtaining very large data sets is prohibitive.

To continue building the cross-disciplinary nature of machine learning applied to microstructure image analysis, we developed an image-driven machine learning approach for the purpose of relating micrographs to annealing parameters (500\degree C for 0-100 hours) for the U-10Mo system. Here, we utilize CNNs for predicting processing from image data, and k-means clustering for image segmentation for the purpose of kinetic modeling. We also demonstrate our approach is viable to analysis of limited original image data sets.

\section{Methods}

\subsection{Experimental Materials and Methods}
\label{Experimental_Methods}

Depleted U alloyed with 10 wt\% Mo (U-10Mo) alloys were studied here in order to perform characterization tasks on U-10Mo alloys with low radioactivity in comparison to prototypic LEU fuel materials which contain 19-20\% \textsuperscript{235}U (relative to all U isotopes).

The samples analyzed in this work include U-10Mo alloys cast and then subjected to a homogenization anneal at 900\degree C for 48 hours, and samples that were cast and homogenized and then annealed at a sub-eutectoid temperature of 500\degree C for 1 to 100 hours. Details on sample fabrication were previously reported in  Reference \cite{Jana2017}. 

It is noted here that although there are several other steps in the fuel fabrication process, the focus of this work was to investigate the specific parameters of time and temperature in a small case study. 

All samples were sectioned, mounted in epoxy and polished using standard metallographic techniques and equipment.  In order to obtain an acceptable surface finish needed for imaging, samples were polished according to the procedure detailed in Reference \cite{Prabhakaran2016}.

\subsection{Image Data}
\label{data_set}

Image data used in this work includes images taken using two scanning electron microscopes (SEMs): a FEI Quanta dual beam Focused Ion Beam/Scanning Electron Microscope (FIB/SEM), and a JEOL JSM-7600F Field Emission SEM. The backscatter electron detector was used for improved atomic number (Z) contrast. All images were taken of metallographically prepared samples by two different microscope operators. Thus the image data used for training and testing was somewhat diverse in terms of resolution, contrast, focus, and magnifications selected. The idea in using a variety of images from different microscopes, but of the same sample was to develop a more robust model that can distinguish between different material processing conditions from different original images. 

All images used in this work are of a depleted U-10Mo alloy fabricated and prepared according to details presented in Section \ref{Experimental_Methods}. Images were taken over a range of magnifications from 250x to 1,000x. Image data was labeled based on processing history, where five material conditions were studied here. Each condition corresponds to a different sub-eutectoid annealing time at 500\degree C: 0 (as-cast and homogenized), 1, 10, 50, and 100 hours. 

Each original image is 1886 by 2048 or 1024 by 1280 pixels including a scale bar region. Prior to training/testing, image data was cropped to remove this scale bar, and to generate a universal image size of 960 by 1280 pixels for training and testing. 140 original images were used for classification (with 16-43 images per class). Prior to segmentation, image pre-processing was performed to remove noise from the original images.  Both median and Gaussian filters were applied to each image, where the median filter size used was 5x5 pixels and the Gaussian sigma was 4. 

\subsection{Data Analysis Approach}

\begin{figure}
  \centering
  \includegraphics[scale=1.0]{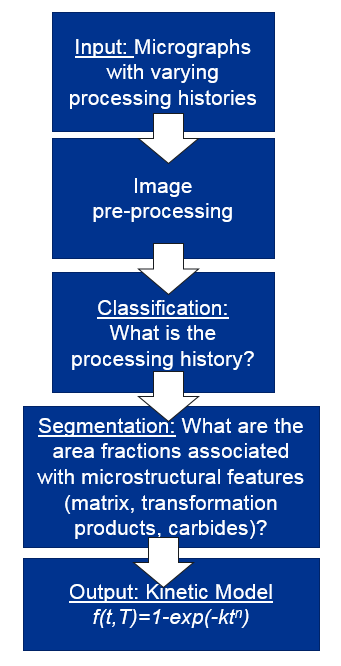}
  \caption{Flowchart of image analysis and kinetic modeling approach.} \label{fig:flowchart}
\end{figure}

The approach developed here builds upon previous work that tested several methods for quantifying microstructure image data and concluded micrographs were best represented by feature vectors extracted using a pre-trained CNN \cite{Chowdhury2016}. In this work, the VGG-19 pre-trained CNN was used \cite{Simonyan2014}, and fine-tuning was performed \cite{FineTuning2016}, which involved only modifying the fifth layer or block in the CNN and freezing all parameters in the first four blocks. Data augmentation method is also used so we have a total of 538 images for training and testing. This method allows for us to use the VGG-19 pre-trained CNN to more effectively represent our microstructure image data, and have been shown to successfully represent image data well versus training a deep CNN from scratch\cite{FineTuning2016}. 
This section details our overall approach (summarized in Figure \ref{fig:flowchart}), including use of machine learning methods for microstructure representation, classification, and segmentation to inform kinetic modeling. 

Data analysis investigated as part of this work involves two main tasks: (1) image classification (where each class corresponds to a different processing history), and (2) segmentation. 

The purpose of image classification is to determine how accurately material processing history can be predicted based on image data alone. For the segmentation task, the purpose is to quantify the microstructure image data and estimate the area fraction corresponding to distinct microstructural features of interest. The micrographs were segmented using the k-means clustering method \cite{Jain2010}, where the features of interest include: the $\gamma$-UMo matrix phase, UC inclusions, and lamellar transformation products of $\alpha$-U + $\gamma$-UMo. Detailed characterization of the as-cast and homogenized, and annealed microstructures can be found elsewhere \cite{Jana2017,Devaraj2018_Acta, Devaraj2018_Scripta}.

We test the accuracy of segmentation by then determining the classification accuracy of segmented images. From segmentation results, the area fraction of lamellar transformation products formed in the discontinuous precipitation reaction were described as a function of processing parameters (time and temperature) by the Johnson-Mehl-Avrami-Kolmogorov (JMAK) equation : 
\begin{equation} \label{KJMA}
f(t,T) =  1-exp(-kt^n )\\
\end{equation}

where $f(t,T)$ is fraction of the lamellar transformation products ($\alpha$U + $\gamma$-UMo), $t$ is annealing time in hours, and $T$ is annealing temperature (500\degree C), $k$ and $n$ are kinetic parameters. $k$ is a time constant, and is inversely proportional to the growth rate of the transformation products. $k$ also depends on volume or nucleation sites in a given volume. $n$ describes dimension of transformation products and is typically an integer value of 1-4 \cite{PE}. The significance of $k$ and $n$ is discussed in additional detail in Section \ref{results}. To extract $k$ and $n$ from segmentation data, the JMAK equation presented in Equation \ref{KJMA} can be rewritten in the following form: 

\begin{equation} \label{KJMA_2}
ln \left(-ln \left(1-f \right) \right) =  nlnt - lnk \\
\end{equation}

To determine values of k and n from the above equation, $ln(-ln(1-f))$ can be plotted versus $ln(t)$. When a line of the form $y=mx+b$ is fit to the data, b is $ln(k)$ and m is $n$. This Avrami analysis has been performed in similar kinetic moedling work reported in References \cite{Jana2017, Semboshi2017}. 

All images are pre-processed in order to first improve contrast between the $\gamma$-UMo matrix phase, the dark carbide inclusions, and the lamellar transformation products. 

For image segmentation, k-means clustering is applied to classify pixels into three clusters based on their grayscale values in each image. The size of each cluster is then calculated, and sorted by mean grayscale of each cluster. As was previously shown in Figure \ref{fig:microstructural_features}, each microstructural feature of interest (matrix, inclusion, lamellar transformation products) has a distinct graycale value, which allows for segmentation based on grayscale to be successful. For each input image the output is three positive numbers that sum to 1, meaning we calculate fraction of each image corresponding to each cluster. Segmented images are classified using a random forest model \cite{Breiman2001}. 

\subsection{Software Specifications and Selection of Model Parameters}
\label{parameters}

All experimentation was carried out with Python version 3.6.7 with the help  of various open-source libraries. The opencv, scipy, skimage, numpy, and sklearn packages (compatible with Python version 3.6.7) were used for training, testing, and validation. An initial investigation of the optimal hyperparamters was made manually by researchers. The Adam optimization strategy was used in training \cite{Kingma2014}.  Mean squared error (MSE) was chosen as the loss for the problem, however, in interpretation of model performance, mean absolute percent error (MAPE) was also calculated to provide a more meaningful metric. Equations for MSE and MAPE are provided here in Equations \ref{MSE_eq} and \ref{MAPE_eq} for completeness. A similar approach was used for displaying training history in prior work \cite{Kautz2019}. Here, $n$ is the number of data points (images), $y_t$ is the labeled image class, and $\hat{y_t}$ is the predicted class.

\begin{equation} \label{MSE_eq}
MSE =  \dfrac{1}{n}\sum\limits_{t=1}^{n}({y_t - \hat{y_t}})^2   \\
\end{equation}

\begin{equation} \label{MAPE_eq}
MAPE =  \dfrac{100\%}{n}\sum\limits_{t=1}^{n}\left |\frac{(y_t - \hat{y_t})  }{y_t}\right|\\
\end{equation}

All relevant model parameters are summarized in Table \ref{tab:parameters}. A GTX 1080 graphics card was used for classification using the pre-trained CNN. 
\begin{table}						
\centering						
\captionsetup{justification=centering}						
\caption{Parameters selected for model specification, compilation, and cross validation.}						
\label{tab:parameters}						
\begin{tabular}{|r|c|c|} \hline						
	&	\textbf{Parameter}	&	\textbf{Value}	\\	\hline
						
\multirow{6}{*}{\textbf{Model Specification}}	&	Optimizer	&	Adam	\\	
	&	Learning Rate	&	5.00E-06	\\	
	&	Body activation	&	ReLU	\\	
	&	Output activation	&	Softmax	\\	
	&	Input dimension	&	(240,320,3)	\\	
	&	Output dimension	&	(5)	\\	\hline
\multirow{3}{*}{\textbf{Compilation}}	&	loss	&	MSE	\\	
	&	optimizer	&	RMSProp	\\	
	&	metric	&	Accuracy, MAPE MSE	\\	\hline
\multirow{5}{*}{\textbf{Cross Validation}}	&	fold	&	5	\\	
	&	training data	&	80\% of augmented images	\\	
	&	testing data	&	20\% of augmented images	\\	
	&	batch size 	&	20	\\	
	&	epochs	&	150	\\	\hline
\end{tabular}						
\end{table}						
				
\begin{figure}
  \centering
  \includegraphics[scale=0.45]{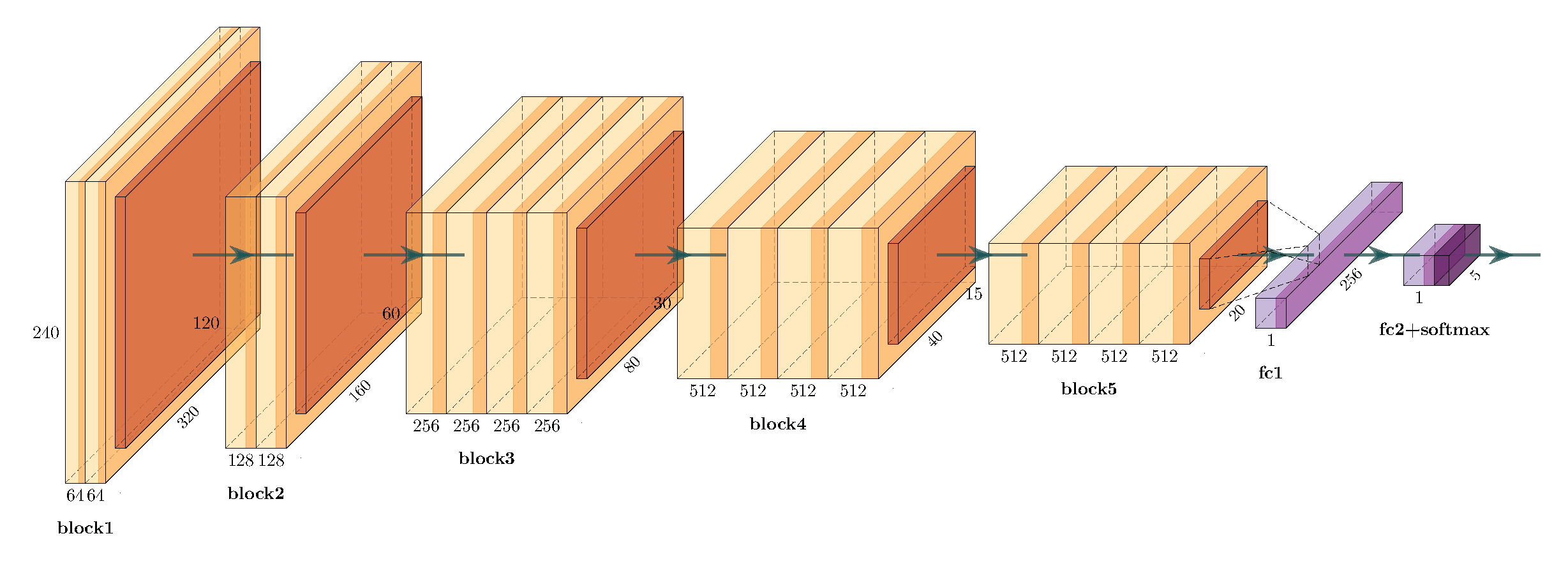}
  \caption{Architecture of the CNN developed for micrograph classification and segmentation. This schematic shows the basic structure of the network.} \label{fig:nn}
\end{figure}

A schematic of the network developed for classification in this work is provided in Figure \ref{fig:nn}. 
\section{Results and Discussion}
\label{results}

\subsection{Microstructure Recognition}
The first step in any microstructure characterization effort is recognition (i.e. identifying the microstructural features of interest, and grouping similar images for subseqent analysis). For the specific task of recognition, we developed a 5-class classification model where microstructure image data is represented by a feature vector extracted by a pre-trained and fine-tuned CNN. A microstructure image of size 960 by 1280 is fed into the model and a feature vector of length 5 is returned. Each element in the feature vector is a floating number between 0 and 1, where the largest element indicates the class the input image belongs to. The architecture of the CNN used in this work is schematically shown in Figure \ref{fig:nn}.

\begin{figure}
  \centering
  \includegraphics[scale=1.15]{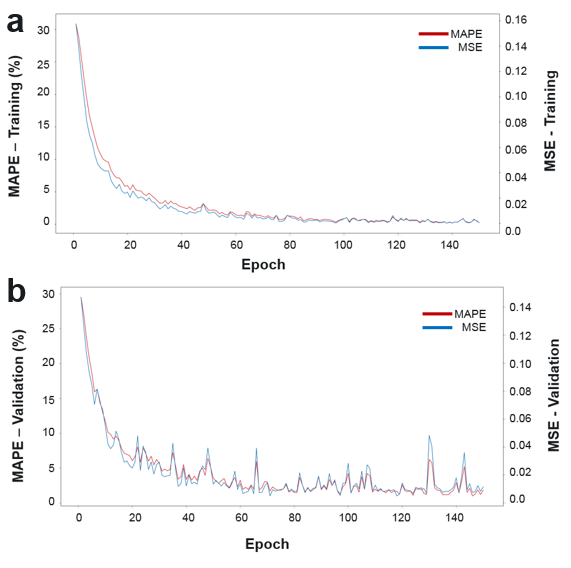}
  \caption{Training history and validation of the CNN developed for micrograph classification. } 
\label{fig:training_history}
\end{figure}

CNN model performance was assessed by investigating model training and validation history. In Figure \ref{fig:training_history}, mean absolute percent error (MAPE) and mean square error (MSE) are plotted versus epoch for both training and validation cases. While MSE is set as the model loss, MAPE is plotted as well as a more meaningful metric for model performance. Here, we can see that for training, MAPE and MSE both decrease and stabilize after approximately 60 epochs, at a MAPE value of less 5\%. This stabilization in model loss indicates that the model has converged and no further performance improvement is obtained with increasing number of epochs. With each epoch, the CNN learns trends in the data set, the MSE decreases and model performance improves. 

A 5-fold cross validation strategy was used to test the trained model performance on never before seen data. After 140 epochs, MAPE is approximately 5\%, which is very low, particularly given the low number of images used in model training.
\subsection{Segmentation}

Next, after images are grouped into 1 of 5 classes, segmentation was performed using the k-means clustering on the grayscale value of each pixel, where k is either 2 or 3, depending on the material condition. For example the as-cast and homogenized sample has only 2 microstructural features of interest (matrix and UC phases), and thus k is 2 for this particular class. However, for images that correspond to as-cast, homogenized, and annealed, there are three microstructural features of interest, thus k is 3 for these material conditions. 

To evaluate the accuracy of segmentation, the segmented images were then classified. When images were classified into 1 of  5 groups (corresponding to the material conditions previously described), a 5-fold cross-validation accuracy 94\% was achieved. This high accuracy indicates that the segmented images can be identified as 1 of the 5 material conditions with a result consistent with classification of the original data. This result also suggests the segmentation of original images represents the original image data well, and the k-means clustering algorithm is a viable method for micrograph segmentation and quantification.

The challenge of image classification based on very limited training data is of significant concern to materials science studies, where there may not be hundreds or thousands of images available, or historic data may need to be mined, in which case additional image data could not be obtained for model development.

\begin{figure}
  \centering
  \includegraphics[scale=1.0]{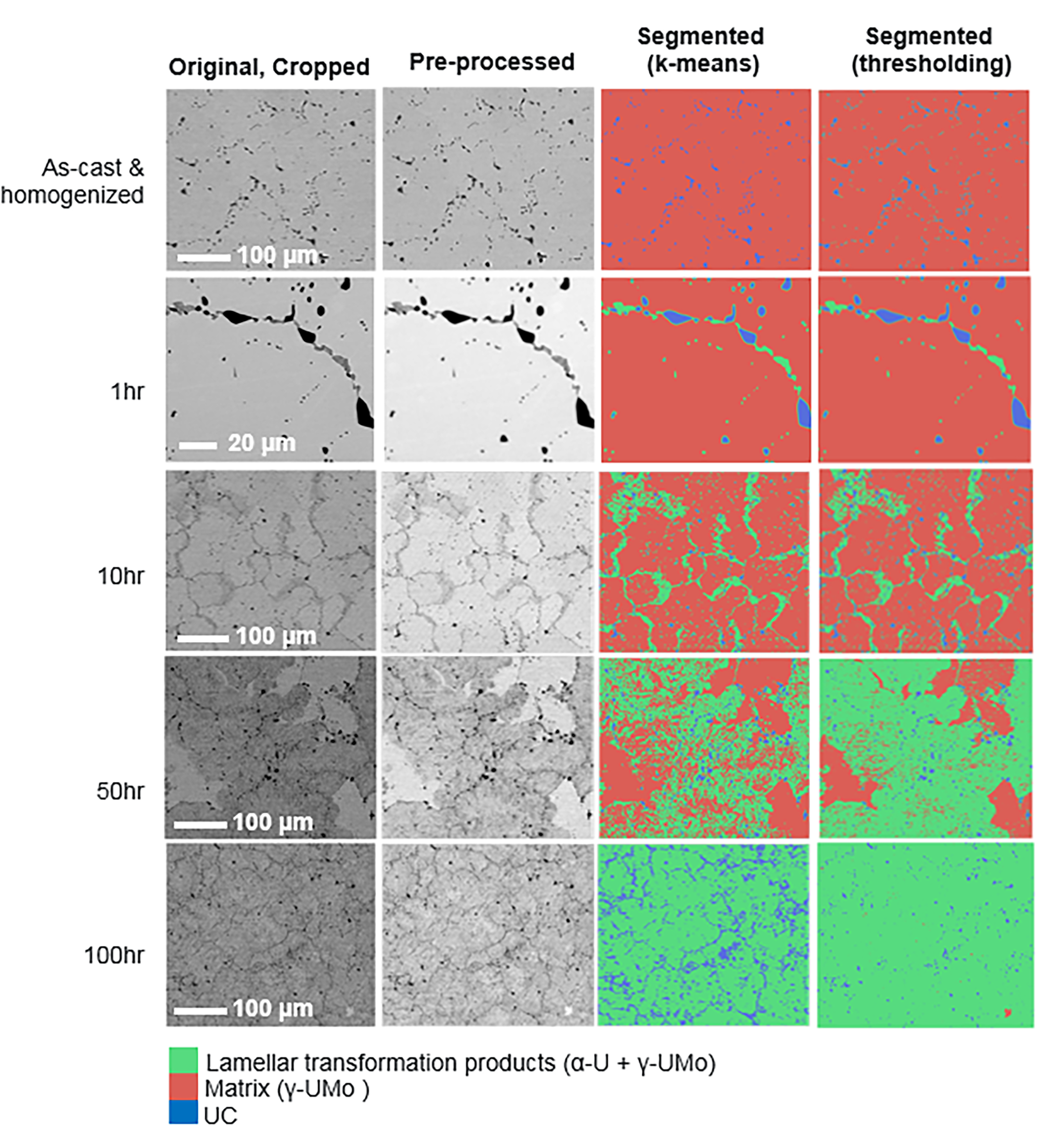}
  \caption{Representative results from image segmentation using the k-means clustering method. Here, a pre-processed image with the scale bar removed from the original image, and enhanced contrast is shown next to the segmented output image. All hours listed correspond to the time at the sub-eutectoid annealing temperature of 500\degree C.} \label{fig:segmentation_results}
\end{figure}

The 5 classes correspond to the previously described processing conditions previously described (as-cast and homogenized, annealed at 500\degree C for 1-100hr). These processing conditions produce microstructures with varying area fractions of the lamellar transformation products, which is accurately captured by image segmentation via k-means clustering. Results from segmentation via manual thresholding and k-means clustering are summarized in Figure \ref{fig:segmentation_results} for all material processing conditions studied in this work.

Based on visual examination of segmentation results, we see that the k-means segmentation accurately segments image data well into UC, matrix, and lamellar transformation products. The segmented images appear very similar between those generated by the k-means algorithm and simple thresholding. In some cases, the segmented images look distinctly different. For example, for that 500\degree C - 50hr anneal example, the segmented image by k-means shows a lower area fraction of the lamellar transformation product than compared to the original image or the thresholded image. This difference could be attributed to the contrast of the original image data. Some regions within the transformed region appear very similar to the matrix grayscale value, and thus the k-means algorithm incorrectly identified some regions as matrix instead of transformation products. Despite this distinct difference, the segmented image via k-means still appears to represent the original image data well. To reduce such errors in area fraction calculation, several images should be segmented and used to generate an area fraction average (as was done in this work). Additionally, the k-means algorithm segmented some grain boundaries as the UC phase, leading to an artificially high UC area fraction for the image corresponding to the 500\degree C - 100hr annealing treatment. Incorrect segmentation of the UC phase inclusions requires additional investigation on how to distinguish between grain boundaries and secondary phases. 
The main advantage to the application of k-means clustering to image segmentation versus thresholding is no requirement for researcher intervention in the image analysis process, and thus the ability to process large amounts of diverse image data obtained from different microscope and from different operators that selected different constrast/brightness, stigmation, and focus settings, and where image resolution may also be different. 

In other microstructures that undergo phase transformations, typically transformation products are identified in micrographs based on their distinctly different grayscale value, thus this method can be easily applied to a different material system for quantifying microstructure image data. While theory regarding nucleation and growth of discontinuous precipitation reaction products have been reviewed in detail \cite{Hillert1972,Williams1981,Hillert1982}, methods for handling image data to quantify the kinetics associated with this particular phase transformation has not been investigated. 
\begin{table}						
\centering						
\captionsetup{justification=centering}				
\caption{Summary of results from area fraction calculations using thresholding (semi-automated), and k-means clustering. All area fractions reported are averages calculated from a minimum of 3 images and corresponding standard deviations are also reported for each phase present: uranium carbide (UC), the lamellar transformation region of alternating $\alpha$ and $\gamma$-UMo phases, and the $\gamma$-UMo matrix.}						
\label{tab:segmentation_results}					
\begin{tabular}{|c|c|c|c|} \hline				
	&		&	\multicolumn{2}{|c|}{Avg Area Fraction $\pm$  Stdev} 							\\ 	\hline
Time at 500\degree C (hr) 	&	Phase 	&	Thresholding			&	k-means 			\\	\hline
0	&	UC	&	0.0531	 $\pm$ 	0.0152	&	0.0327	 $\pm$ 	0.0049	\\	
	&	$\alpha$-U + $\gamma$-UMo	&	0.0000	  $\pm$  	0.0000	&	0.0000	  $\pm$  	0.0000	\\	
	&	$\gamma$-UMo matrix	&	0.9469	 $\pm$   	0.0152	&	0.9673	 $\pm$   	0.0049	\\	\hline
1	&	UC	&	0.0225	 $\pm$ 	0.0027	&	0.0142	 $\pm$ 	0.0030	\\	
	&	$\alpha$-U + $\gamma$-UMo	&	0.0420	  $\pm$  	0.0104	&	0.0308	  $\pm$  	0.0044	\\	
	&	$\gamma$-UMo matrix	&	0.9356	 $\pm$   	0.0131	&	0.9549	 $\pm$   	0.0061	\\	\hline
10	&	UC	&	0.0495	 $\pm$ 	0.0100	&	0.0206	 $\pm$ 	0.0029	\\	
	&	$\alpha$-U + $\gamma$-UMo	&	0.1852	  $\pm$  	0.0475	&	0.1536	  $\pm$  	0.0671	\\	
	&	$\gamma$-UMo matrix	&	0.7653	 $\pm$   	0.0564	&	0.8258	 $\pm$   	0.0687	\\	\hline
50	&	UC	&	0.1174	 $\pm$ 	0.0140	&	0.0467	 $\pm$ 	0.0461	\\	
	&	$\alpha$-U + $\gamma$-UMo	&	0.6089	  $\pm$  	0.0162	&	0.5144	  $\pm$  	0.0695	\\	
	&	$\gamma$-UMo matrix	&	0.2737	 $\pm$   	0.0134	&	0.4389	 $\pm$   	0.0610	\\	\hline
100	&	UC	&	0.0362	 $\pm$ 	0.0124	&	0.1545	 $\pm$   	0.0898	\\	
	&	$\alpha$-U + $\gamma$-UMo	&	0.9154	  $\pm$  	0.0193	&	0.8455	  $\pm$  	0.0898	\\	
	&	$\gamma$-UMo matrix	&	0.0484	 $\pm$   	0.0069	&	0.0000	 $\pm$ 	0.0000	\\	\hline

\end{tabular}						
\end{table}	
\begin{figure}
  \centering
  \includegraphics[scale=0.85]{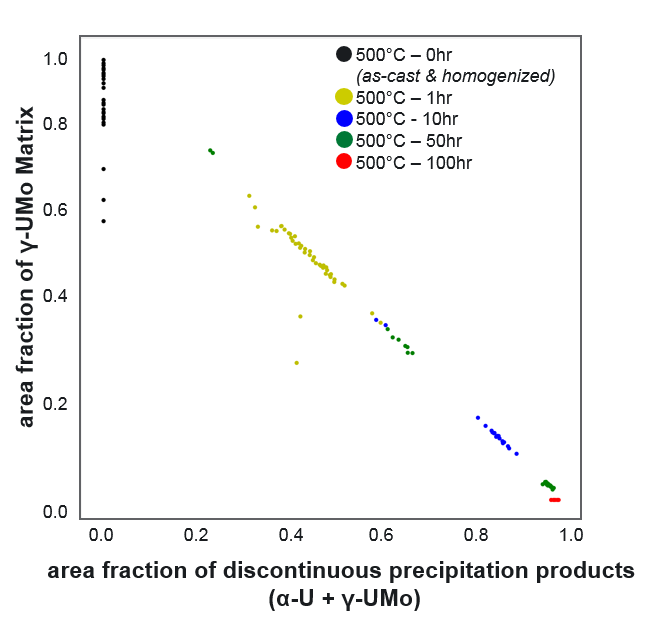}
  \caption{Visualization of image segmentation results, where area fraction of the matrix is plotted versus area fraction of the lamellar transformation products (formed by the discontinuous precipitation reaction). Each material processing condition (sub-eutectoid annealing temperature and time) is plotted in a different color, as indicated in the legend.} \label{fig:segmentation_visualization}
\end{figure}
Figure \ref{fig:segmentation_visualization} clearly shows that as the area fraction of the $\gamma$-UMo matrix decreases, the area fraction of the transformation products increases, which we observe when visually examining micrographs.  This trend is consistent with the understanding of this particular transformation in which the lamellar products consume $\gamma$-UMo grains. The area fraction of the UC inclusions is not included in this plot, because this value was consistent for each group of images analyzed, as previously summarized in Table \ref{tab:segmentation_results}. While segmentation results can be visually seen by comparing each original image to the corresponding segmented image, the plot provided in Figure  \ref{fig:segmentation_visualization} provides a different way of investigating segmentation results. The data points lying along a straight line indicates that the calculated area fractions for the transformed region (needed for studying kinetics of phase transformation in this system) are accurate, because if the UC area fraction is consistent, the remaining image consists of lamellae and matrix, and the totals of these two area fractions should sum to close to 1.0 (less the area fraction of carbide inclusions). Thus these two features are linearly related, which is confirmed in the Figure \ref{fig:segmentation_visualization} plot. 

\subsection{Kinetic Modeling}

\begin{figure}
  \centering
  \includegraphics[scale=0.75]{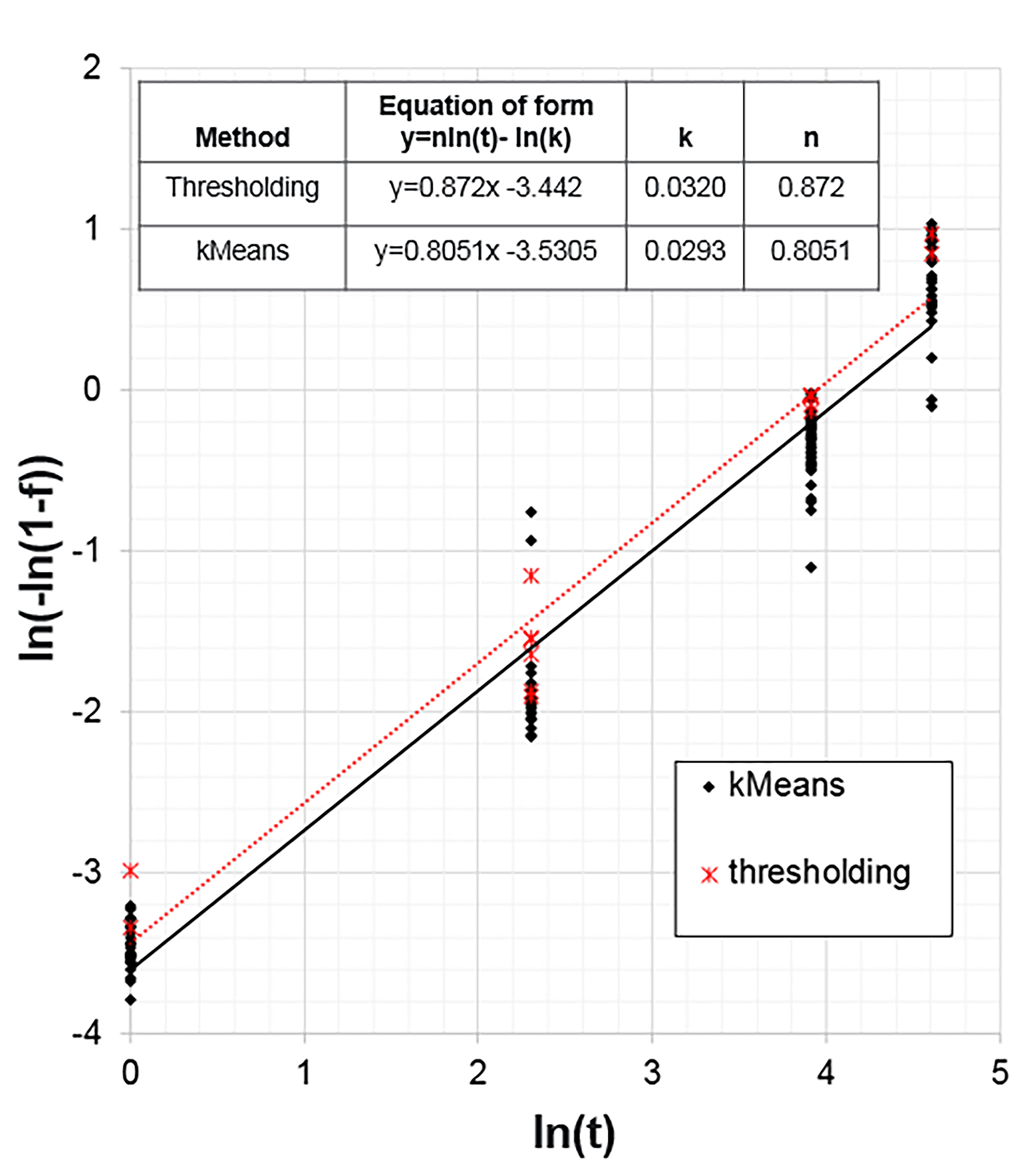}
  \caption{Avrami analysis of data from as-cast, homogenized and heat-treated samples at 500\degree C for 1-100 hours. Here, area fraction of lamellar transformation products is variable $f$, and $t$ is time in hours. Results from thresholding and k-means clustering are plotted together here. Kinetic parameters are summarized for both methods of microstructure quantification.} \label{fig:KJMA}
\end{figure}

From segmentation results and with the assumption that the discontinuous precipitation reaction can be described by the JMAK equation, we can plot our segmented results as a function of time, as shown in Figure \ref{fig:KJMA}. By fitting a line to data obtained from segmentation, the kinetic parameters of k and n in the JMAK equation were determined to be 0.0293 and 0.8051, respectively, based on segmentation via k-means, and 0.0320 and 0.872 for thresholding. 

Results from thresholding (in a semi-automated methods, as previously described) and k-means segmentation are plotted together in order to compare results from each approach. The difference in measured area fraction of the discontinuous precipitation products varies between 0.01112 - 0.094478 when comparing k-means versus manual thresholding. In all cases, the k-means clustering algorithm yields a lower calculated area fraction of the discontinuous precipitation products in comparison to thresholding. Based on the results summarized in Figure \ref{fig:KJMA}, similar k and n values can be obtained from these two methods, and thus in terms of kinetic modeling, the new approach for image analysis used here is not distinctly different than previously used manual thresholding, as reported in Reference \cite{Jana2017}. However, our approach does allow for a more standardized analysis of a more image data, without extensive domain knowledge or intervention from the researcher. 

As previously discussed in Reference \cite{Jana2017}, the value of n is related to nucleation and growth, whereas k is related to reaction rate. Typical values of n are from 1 to 4. A value of ~1 for n indicates diffusion controlled growth of the transformation products of interest. Here, values of n close to 1 were calculated based on segmentation results, suggesting the discontinuous precipitation reaction is diffusion-controlled, and nucleation grain boundaries \cite{Kolmogorov1991,Chai2015}. The calculated n-values here are consistent with previously reported results in Reference \cite{Jana2017}. Any small differences in the kinetic parameters calculated in this work versus that reported in Reference \cite{Jana2017} can be attributed to different image analysis approaches, where in prior work, the UC phase was not separately segmented from the transformed region, and fewer number of images were used in analysis. 

A key conclusion from this Avrami analysis is that the image analysis/quantification approach developed in this work successfully describes image data well for the purpose of kinetic modeling and developing microtructure-processing relationships. The approach developed here can easily be transferred to different images and processing conditions within this system, or adapted for a system in which phase transformation kinetics can be better understood by leveraging microstructure image data.  

\section{Limitations}
\label{limitations}

Limitations associated with the data analysis approach presented here are generally related to reliance on domain knowledge, the amount of original image data available, and visualization of results.

Image data labeling was performed using based on domain knowledge of SEM imaging and the U-10Mo material system. Images were labeled according to the sub-eutectoid annealing time at 500\degree C. Low-level pixel processing was performed in order to determine the ground truth of area fraction measurements used to compare to results from deep learning methods. Additionally, 2D image data is used in this work to quantify amount of phase transformation product, which is a 3D object. Thus 2D image data is an inherently limited representation of material microstructure. Obtaining 3D information on microstructures is possible via serial sectioning and data reconstruction methods, however these methods are much more time consuming in comparison to simpler imaging only via SEM or optical methods. It is also of primary interest in this work to investigate microstructure quantification/characterization methods that can be applied to existing image data sets in order to leverage the wealth of image data in materials science communities. Thus, the data analysis approach presented here is limited to using area fraction as a proxy for volume fraction, and to limited image data sets for training/testing machine learning models. With increased number of raw image data per class, classification accuracies would likely improve. Additional heat treatment times at the 500\degree C annealing temperature would also be helpful in developing a more representative kinetic model for this particular system. With additional heat treatment times, calculated values of $k$ and $n$ in the JMAK equation may be more accurate. 

Lastly, a known disadvantage or limitation of using CNNs for microstructure representation is that it can be difficult to interpret intermediate features within the neural network architecture, as noted by Kono \textit{et al} in related work \cite{Kondo2017}.  To allow for interpretation of features extracted by the CNN, activation maps can be utilized as was presented by Ling \textit{et al}.

\section{Conclusions}
\label{conclusions}

Micrograph quantification is an essential component of to understanding microstructure-processing-property-performance relationships in a wide range of material systems. 
A more efficient and less biased methodology would enable more rapid discoveries in the field of materials science, and allow for leveraging the wealth of knowledge available in micrographs. Here, we address this need for improved micrograph image analysis approaches and tools by developing an image-driven machine learning approach to microstructure quantification which involves the following tasks: (1) prediction of material processing history based on image data (classification), (2) calculation of area fraction of phases present in the micrographs (segmentation), and (3) kinetic modeling from segmentation results. This approach was developed in a case study on the U-10Mo system, which experiences a discontinuous precipitation reaction during sub-eutectoid annealing at 500\degree C between 1-100 hours. We used a total of 140 original BSE-SEM micrographs for model development and analysis. 

The purpose of this work is to enable the development of microstructure-processing relationships in metal alloy systems in which routine image analysis, particularly image segmentation, is required. The approach developed and presented here is more efficient and less biased than more commonly used manual methods of low-level pixel processing.  

The following conclusions and insights can be drawn from this work:
\begin{itemize}
    \item Number of original images, and their quality can significantly impact classification and segmentation results. 
    
    \item A pre-trained and fine-tuned CNN (here, VGG-19) represents microstructure image data well for the purpose of micrograph recognition in a multi-class problem. A classification  accuracy of 94\% (for 5-fold cross validation) was achieved despite very limited image data available for model training and testing (less than 45 original images per class for training/testing).
    
    \item Segmentation using the k-means clustering algorithm yielded results (area fraction of different phases) that agreed well with manually annotated images. 
    
    \item Classification of segmented images yields an accuracy of 94\% (same accuracy obtained for original image data classification) suggesting that the k-means clustering algorithm accurately represents original data.
    
    \item Kinetic modeling results agree well with previously reported data that was developed using a more manual approach to image analysis. The application of CNNs to classification, and k-means clustering for segmentation allows for researchers to leverage information in small image data sets for understanding trends in microstructural evolution.

\end{itemize}

\subsection{Future Work}
Future work includes studying larger image data sets, and studying both thermal and mechanical processing steps on microstructural evolution.   Additionally, segmentation via CNNs could be investigated and visualization of texture features could also be performed in order to correlate with kinetics of the discontinuous precipitation reaction. This approach to visualization of texture features is detailed in Reference \cite{Ling2017}. Further, the generalizability of the methods presented here should be investigated further by applying the workflow presented here to kinetic modeling in a different system other than U-10Mo. Our approach has the potential to generalize to multiple material systems, since several others works (previously discussed in Section \ref{intro}) also apply CNNs to microstructure image data representation, indicating that CNNs are a flexible approach for studying material science micrographs. 
\section*{Acknowledgments}
\label{acknowledgments}

This work was conducted at Pacific Northwest National Laboratory operated by Battelle for the United States Department of Energy. This work was funded by the U.S. Department of Energy National Nuclear Security Administration's Office of Material Management and Minimization and performed at Pacific Northwest National Laboratory (PNNL) under contract DE-AC05-76RL01830. The authors acknowledge Dr. Aritra Chowdhury for his work in image classification that was essential in paving the way for the work presented here. The authors also wish to acknowledge personnel at PNNL for experimental work done to generate data used in this work, in particular the following individuals: Shelly Carlson (PNNL) for metallographic sample preparation for subsequent imaging, and Alan Schemer-Kohrn (PNNL) for expertise in electron microscopy and image acquisition of image data used in this work.

\section*{Contributions}
D.J.L. and B.Y. conceptualized the data analysis approach presented here. E.J.K performed all image analysis and kinetic modeling, and lead manuscript writing. W.M. performed machine learning methods development. V.J. directed sample preparation and research efforts presented here. S.J. and A.D. contributed to discussions of results and significance to phase transformations in the U-Mo system. All authors contributed to manuscript preparation.
\section*{Data Availability}
The raw/processed data required to reproduce these findings cannot be shared at this time due to technical or time limitations.



\section*{References}

\bibliography{CompVision_EJK}

\end{document}